# A Madelung-Buckingham Model for the Variation of the Cubic Lattice Constant of $Li_xMn_2O_4$ during the charge/discharge of the Lithium-ion Battery

**Daniel Sherwood and Bosco Emmanuel**[*]

Modelling & Simulation Group

Central Electrochemical Research Institute, Karaikudi 630 006, Tamil Nadu, India

**Abstract** $Li_xMn_2O_4$ is an important cathode material for the Li-ion battery. During charging, the stoichiometry x varies continuously from 1 to zero and on discharging it varies from zero to one. The cubic lattice constant 'a' of $Li_xMn_2O_4$ depends on the value of x. The variation of 'a' with x has important consequences for battery performance. In this paper, we use a Madelung-Buckingham model to study this variation and compare the results with experimental data on $Li_xMn_2O_4$.

## Introduction

Research on Lithium-ion Batteries is being actively pursued all over the globe for their applications in several areas including Electric Vehicles[1,2]. Lithium manganate ($Li_xMn_2O_4$) is an important cathode material for this class of batteries[3]. Though it is less expensive and more eco-friendly in comparison to other cathode materials such as the Lithium cobaltate ($Li_xCoO_2$), $Li_xMn_2O_4$ suffers from capacity losses[4]. There is an extensive research to study the origin of these losses and to evolve means to mitigate them. Capacity fade in $Li_xMn_2O_4$ are of two kinds: reversible capacity loss and irreversible capacity loss. The reversible capacity loss arises from the low mobility and hence long diffusion path lengths for Li-ion transport in the lithium manganate crystallites. Reversible capacity loss, can, in principle, be minimized at smaller currents and over a larger discharge time. Irreversible capacity loss is related to : (i) manganese dissolution from $Li_xMn_2O_4$ cathode into the battery electrolyte[5] and (ii) volume changes in the host lattice upon charge/discharge. It is the capacity loss under the category (ii) which the present paper is concerned with. This irreversible capacity loss arises thus: the lithium battery cathode, of our interest here, is a composite film consisting of $Li_xMn_2O_4$ interspersed with carbon powder. $Li_xMn_2O_4$ is a poor electrical conductor and the carbon particles which sit in between the manganate crystallites help to improve the electrical conductivity of the composite film[6]. However, on charging, lithium is de-intercalated from the $Li_xMn_2O_4$ particles leading to a decrease in their crystal volume and, on discharging, lithium is intercalated back and the crystal volume increases. Hence, when the battery is repeatedly charged and discharged, the lithium manganate particles expand and contract resulting in an irreversible loss of inter-particle contacts and hence an increased capacity loss of the composite cathode film.

$Li_xMn_2O_4$ has a cubic structure of Fd3m symmetry in which the $Li^+$ and $Mn^{3+}/Mn^{4+}$ ions are located in the 8a tetrahedral and 16d octahedral sites, respectively, in a cubic close-packed array of $O^{2-}$ ions, which occupy the 32e sites[7,8]. The 8a lithium sites together with the vacant 16c octahedral sites provide the three-dimensional channels (8a–16c-8a) for lithium intercalation and deintercalation. Lithium can be removed from the 8a sites at 4.1V vs. $Li^+ / Li$, the complete delithiation yielding $\lambda - MnO_2$. The cyclability of $Li_xMn_2O_4$ is determined by the structural integrity of the host lattice during the intercalation-deintercalation process[9]. The charge-discharge process in the 4V region is accompanied by a 7.6% volume change in the unit cell. However, the volume change is so gradual and isotropic that the cubic symmetry of the material is usually maintained[10]. Nevertheless repeated cycling especially at elevated temperatures ( > 55º C) results in a capacity fade [11-13].

Minimizing the crystal volume change on charge/discharge will, therefore, greatly enhance the charge-discharge performance of the cathode. In this paper, we use a Madelung-Buckingham model to theoretically simulate the variation of the cubic lattice constant of $Li_xMn_2O_4$ during charge/discharge. Although we have applied the model to $Li_xMn_2O_4$ in this paper, it can be applied to any ionic crystal amenable to a Madelung-Buckingham-type description. Of particular interest are a host of oxides employed as cathodes in lithium-ion batteries. We propose that the model will be useful in screening them for their susceptibility to irreversible capacity loss due to crystal volume changes on charge/discharge.

**Madelung-Buckingham Model**

The lattice constant 'a' of the cubic crystal $Li_xMn_2O_4$ is that value of 'a' which minimizes the crystal energy. The energy of an ionic crystal is composed of 3 major energy terms as in eq 1.

$$\text{Energy} = \text{Madelung} + \text{Short-range electron-electron repulsion}$$
$$\text{(long-range coulombic)}$$
$$+ \text{van der Waals'} \quad (1)$$

where the Madelung and the van der Waal's components are attractive forces. The short-range electron-electron repulsion and the van der Waal's forces will be modeled by the Buckingham Potential:

$$A_{ij} \exp(-r_{ij}/\rho_{ij}) - C_{ij}/r_{ij}^6 \quad (2)$$

where the $A_{ij}$, $\rho_{ij}$ and $C_{ij}$ are the relevant Buckingham parameters for the ion-pair (i, j) and $r_{ij}$ is the distance of separation.

The Madelung energies are well known in Solid State Physics and Chemistry [14,15]. The Ewald method is a powerful tool to evaluate Madelung energy [16]. However, unlike the Madelung energy computations for conventional crystals like NaCl, we need to evaluate, for our work, the Madelung energy for values of x in the range 0 to 1 in $Li_xMn_2O_4$. x is an additional variable which should be properly incorporated within an Ewald procedure. This leads to the following expression[#] for the Madelung energy of $Li_xMn_2O_4$ as a function of x.

$$E_M(x) = \tfrac{1}{2}\left[\sum_{g \neq 0}(SS^*)f(g) + \left\{\sum_{i=1}^{N}\lambda^2(i)\right\}F(G) + \sum_{j=1}^{N}\lambda(j)\sum_{i \neq j}^{N}\lambda(i)\overline{F}(G,r_i^j)\right] \quad (3)$$

where

- $g$      denotes the reciprocal lattice vector

- $l$      denotes the real-space lattice vector

- $N$      is the number of ions in the crystal basis

- $S$      $\sum_{i=1}^{N}\lambda(i)\exp(ig.r_i)$

- $S^*$      the complex conjugate of S

- $r_i^j$      $r_i - r_j$

- $r_i$      the atomic position of the $i^{th}$ ion in the basis

$f(g) = (\pi/v_c)\,(1/G^2)\exp\text{-}(g^2/4G^2)/(g^2/4G^2)$

$F(G) = \sum_{l \neq 0}(1/|l|)\,erfc\{G|l|\} - 2G/\sqrt{\Pi}$

$\overline{F}(G,r_i^j) = \sum_{l}(1/|l + r_i^j|)\,erfc\{G|l + r_i^j|\}$

and    $v_c$ is the unit cell volume.

---

\# The details of the derivation are provided in the Appendix.

In eq 3, G is a scalar parameter which is adjusted for fast convergence of the infinite sum. The variable stoichiometry x enters through the $\lambda(i)$'s which allow for both the charge and the partial occupancy at the $i^{th}$ site in the basis. For $Li_xMn_2O_4$, the primitive basis has two Li sites (8a), four Mn sites (16d) and eight O sites (32e). The corresponding $\lambda(i)$'s are :

$$\lambda(1) = \lambda(2) = x, \lambda(3) = \lambda(4) = \lambda(5) = \lambda(6) = 4 - x/2$$
$$\lambda(7) = \lambda(8) = ......... = \lambda(14) = -2$$
(4)

The contribution arising from the last two terms in eq. 1 can be written as

$$E_B(x) = \sum_{(i,j)} N_{ij} f_i f_j \left\{ A_{ij} \exp(-r_{ij}/\rho_{ij}) - C_{ij}/r_{ij}^6 \right\}$$
(5)

where $(i, j)$ runs over the nearest neighbour and the next nearest neighbour ion-pairs and $f_i$, $f_j$ are the $x$-dependent occupancies at sites i and j respectively. $N_{ij}$ are the number of pairs of the type $(i, j)$ per formula unit.

**Computational Details**

$Li_xMn_2O_4$ is a cubic spinel with space group Fd3m. The primitive basis has two lithium ions, four manganese ions and eight oxide ions. The oxide ion valence can be considered fixed at –2 and lithium valence at +1. This crystal is a mixed-valent compound with respect to the oxidation state of the manganese ion. When the stoichiometry x of the spinel varies from 0 to 1 the valence of the manganese ion continuously varies from 4+ to a mixed valent state of 50 % 4+ and 50 % 3+.

For the computation of the Madelung part of the total energy, the number of ions in the primitive basis (which is 14 for $LiMn_2O_4$), the atomic positions of these fourteen ions and the $\lambda(i)$'s enter as inputs into eq 3. A value of unity for the convergence factor G was found to be optimal for the summations. A comparison with known Madelung constants of conventional crystals such as NaCl, CsCl and ZnS showed that the computational accuracy was at least up to 5 decimal places for a value of G = 1 and for grids of size (10x10x10) both in the real and reciprocal spaces.

It may further be noted that, for the cubic system $Li_xMn_2O_4$, the unit-cell constant 'a' can be chosen as a convenient length scale and, hence, the Madelung energy can be expressed as

$$E_M(x) = -\frac{f(x)}{a} \qquad (6)$$

where the function $f(x)$ depends only on the stoichiometry x and is independent of 'a'.

The last two terms in the energy expression eq 1 were computed using the Buckingham potential as in eq 5. Ammundsen and co-workers[17] have computed the Buckingham parameters $A_{ij}$, $\rho_{ij}$ and $C_{ij}$ appearing in eq 5 from the vibrational spectra of $LiMn_2O_4$. These and other parameters used in the computation of eq 5 are listed in Table 1. Note also that the inter-ionic distances $r_{ij}$ can also be scaled by the lattice constant 'a' and written as $r_{ij} = \left(\frac{r_{ij}}{a}\right) * a$, where $\left(\frac{r_{ij}}{a}\right)$ is a non-dimensional constant, denoted $\overline{r}_{ij}$ in the Table 1.

**TABLE 1: The Parameters used for the Madelung Buckingham Computation**

| ion-pair $(i,j)$ | $N_{ij}$ | $f_i$ | $f_j$ | $A_{ij}$ (eV) | $\rho_{ij}$ (Å) | $C_{ij}$ (eVÅ$^6$) | $\overline{r}_{ij}$ |
|---|---|---|---|---|---|---|---|
| $O^{2-} \ldots O^{2-}$ | 24 | 1 | 1 | 22764.3 | 0.149 | 43 | 0.3361 |
| $Li^+ \ldots O^{2-}$ | 4 | x | 1 | 426.48 | 0.300 | 0.0 | 0.2376 |
| $Mn^{3+} \ldots O^{2-}$ | 12 | x/2 | 1 | 1267.5 | 0.324 | 0.0 | 0.2381 |
| $Mn^{4+} \ldots O^{2-}$ | 12 | (1-x/2) | 1 | 1345.15 | 0.324 | 0.0 | 0.2381 |

The ion-pairs in Table 1 are the nearest neighbour pairs. The second nearest neighbour interaction was found significant, in addition to the first, for the $O^{2-} \ldots O^{2-}$ pair and was included in the computation, while for the other ion-pairs the nearest neighbour interactions were adequate.

**Results and Discussion**

Combining eqs 1, 5 and 6, we get

$$E(x,a) = E_M(x,a) + E_B(x,a)$$

$$= -\frac{f(x)}{a} + \sum_{(i,j)} N_{ij} f_i(x) f_j(x) \left\{ A_{ij} \exp(-\bar{r}_{ij} a / \rho_{ij}) - (C_{ij} / \bar{r}_{ij}^{\,6}) a^{-6} \right\} \quad (7)$$

It is to be noted that the total energy E depends on the lattice constant 'a' in addition to the stoichiometry index x.

For each value of x ( in the range between 0 and 1 ), the value of 'a' can be obtained by minimizing E(x,a) w.r.t 'a'.

i.e. $\quad \dfrac{\partial E(x,a)}{\partial a} = 0$

Results of this minimization for several values of x are shown in figures 1 and 2.

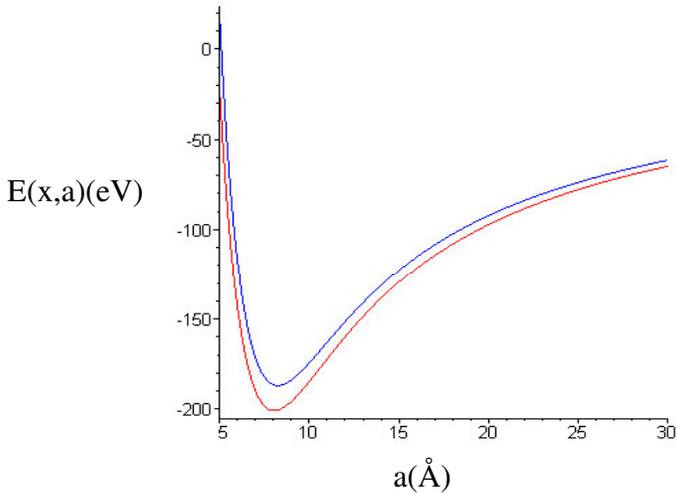
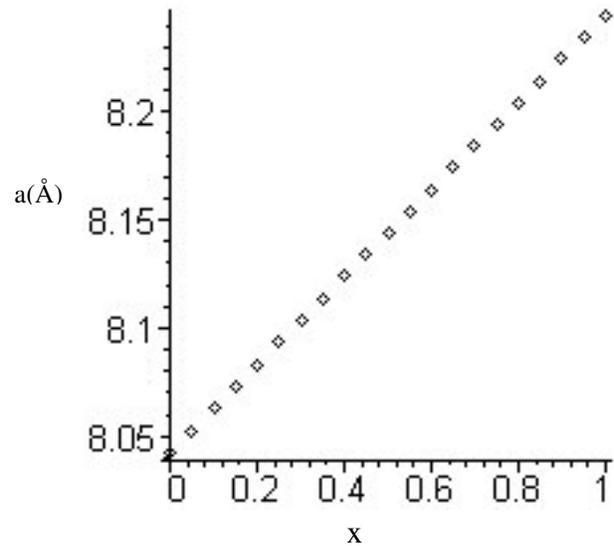

Figure 1: Energy E(x,a) plotted against a, red line for x=0 and the blue line for x=1

Figure 2: The Lattice constant 'a' in Å versus the stoichiometry x

The predicted variation of the lattice constant 'a' with x is in excellent agreement with experimental data reported in the literature for $Li_xMn_2O_4$ [see page 37 of ref. 1] . Moreover, it is interesting to note that the variation of 'a' is linear in x, despite the non-linearities manifest in eq 7.

Doped analogues $Li_xM_yMn_{2-y}O_4$ of $Li_xMn_2O_4$, where M = Fe, Co, Ni, Cr etc., are also of interest[5] for lithium-ion batteries. The methodology presented in this paper is applicable to these oxides as well. However, at present the Buckingham parameters for these dopant species are not available. One way to obtain these parameters is from the vibrational spectral data, as done by Ammundsen etal[17]. An alternative approach is to obtain these from data such as the compressibility, the Gruneisen parameter etc[18]. These latter data are also not yet available for these systems.

**Conclusions**

It is pertinent to note that Ceder and co-workers[19,20] have carried out interesting quantum ab initio calculations on systems of the general formula $Li_xMO_2$ ( M = Co, Ni, Cu, Mn etc). They computed total energies from which the battery voltages were derived. A typical calculation of the total energy requires nearly 1 hour on a Cray C90 supercomputer. In addition, with the currently available computational resources, the x = 0 and x = 1 stoichiometries only are amenable for a quantum computation[20]. For x $\neq$ 0 or 1, the structures are non-periodic on the atomic scale ( due to the disorder on the lithium sites) or have a large periodicity ( if lithium orders into super-structures). In comparison with these quantum simulations, the classical simulations presented here are not computationally very demanding, and all values of x can be treated with equal ease. It takes nearly 8 hours on a 1.8 GHz Pentium IV PC for computing the results for the full range of x from 0 to 1. Though, the superiority of the ab initio methods must be admitted, currently available computational resources will not permit the quantum calculation of crystal volume changes for $Li_xMn_2O_4$. Hence the classical method described in this paper is particularly attractive.

**Acknowledgements**   The authors thank Dr.T.Prem Kumar ( Li-ion Battery Group,CECRI) for reading and improving the manuscript.

**Appendix**

**Computation of $E_M$**

In this Appendix, Ewald's technique is applied to compute the long-range electrostatic interactions in ionic crystals of variable stoichiometries and mixed valencies. Any ionic crystal may be specified by giving its crystallographic space group, the unit cell parameters (corresponding to the primitive, conventional or super cells) and the corresponding basis (consisting of a set of ions). The electrostatic energy of ionic crystals is usually expressed as a sum of pair wise coulombic terms given by

$$E_M = \sum_{(i,j)} \frac{z_i z_j}{r_{ij}} \tag{i}$$

where $z_i$ and $z_j$ are the valencies of the i$^{th}$ and j$^{th}$ ion and $r_{ij}$ is the interionic distance. The sum runs over all ion pairs. In order to apply Ewald's method for crystals of variable stoichiometry and mixed valency, the above sum is expressed in terms of contributions arising from several sublattices present in the crystal so that the stoichiometry and the valency can be tuned in each sublattice. Hence the appropriate form for the energy will be

$$E_M = \frac{1}{2} \sum_{i_{ref}=1}^{N} E_{iref} \tag{ii}$$

where N is the number of ions in the basis and is also the number of sub-lattices into which the crystal can be split. The factor 1/2 removes the double counting of the pair interaction.

$E_{iref}$ is the energy of interaction of any chosen reference ion with its own Bravais relatives[#] and with other ions in the basis and their Bravais relatives.

_________________________________________________________________

\# Bravais relatives of a given ion are here defined as the set of ions generated by Bravais translations acting on the chosen ion.

Let $\mathbf{r}_i = [x(i), y(i), z(i)]$     $i = 1 \rightarrow N$

denote the atomic positions of the $i^{th}$ ion in the basis and $\lambda(i)$  $i = 1 \rightarrow N$ denote the effective charge at the $i^{th}$ ion of the basis. Shift the origin of the co-ordinates (0,0,0) so that $\mathbf{r}_{iref} = (0,0,0)$. In this co-ordinate system

$\mathbf{r}_i [x(i)-x(i_{ref}), y(i)-y(i_{ref}), z(i)-z(i_{ref})] = \mathbf{r}_i'$

Now the interaction energy $E_{iref}$ can be written as

$$E_{iref} = \sum_{l \neq 0} [\lambda^2(i_{ref}) / |l|] + \sum_{i \neq i_{ref}}^{N} \sum_{l} \lambda(i_{ref}) \lambda(i) / |l + \mathbf{r}_i'|$$

$$= \lambda^2(i_{ref}) \sum_{l \neq 0} 1/|l| + \lambda(i_{ref}) \sum_{i \neq i_{ref}}^{N} \lambda(i) \sum_{l} 1/|l + \mathbf{r}_i'| \qquad (iii)$$

$$E_{iref} = \lambda(i_{ref}) [(\sum_{l \neq 0} \lambda(i_{ref}) / |l|) + (\sum_{i \neq i_{ref}}^{N} \lambda(i) \sum_{l} 1/|l + \mathbf{r}_i'|)] \qquad (iv)$$

In the above equations $l$ is the Bravais translation vector given by

$l = l_1 \mathbf{a} + l_2 \mathbf{b} + l_3 \mathbf{c}$

where the vectors **a, b** and **c** depend on the type of unit cell chosen.

Using Ewald's transformation the summations appearing in equation (iv) can be expressed as

$$\sum_{l \neq 0} 1/|l| = \sum_{g} f(\mathbf{g}) + F(G) \qquad (v)$$

$$\sum_{l} 1/|l + \mathbf{r}_i'| = \sum_{g} \exp(-i\mathbf{g} \cdot \mathbf{r}_i') f(\mathbf{g}) + \bar{F}(G, \mathbf{r}_i') \qquad (vi)$$

where $f(\mathbf{g}) = (\pi/v_c) \cdot (1/G^2) \cdot \exp{-(\mathbf{g}^2/4G^2)} / (\mathbf{g}^2/4G^2)$ \qquad (vii)

$$F(G) = \sum_{l \neq 0} (1/|l|) \, \text{erfc} \{G.|l|\} - 2G/\sqrt{\pi} \qquad (viii)$$

$$\text{and } \overline{F}(G, \mathbf{r}_i') = \sum_{l} (1/|l + \mathbf{r}_i'|) \, \text{erfc} \{G.|l + \mathbf{r}_i'|\} \qquad (ix)$$

In the above equations, **G** is a variable scalar parameter which is adjusted for fast convergence of the infinite sum, **g** is the reciprocal lattice vector given by **g** = h**A**+k**B**+l**C** where vectors **A, B, C** are obtained from the vectors **a, b** and **c** by the usual transformations. $v_c$ is the unit cell volume given by $v_c = |\mathbf{a} \times \mathbf{b} \cdot \mathbf{c}|$.

$E_{i\,ref}$ may now be written as

$$E_{i\,ref} = \lambda(i_{ref}).[\,\lambda(i_{ref}).\sum_{g} f(g) + \lambda(i_{ref}).F(G) + \sum_{g} \{\sum_{i \neq iref}^{N} \lambda(i) \exp{-ig \cdot \mathbf{r}_i'}\} f(g)$$

$$+ \sum_{i \neq iref}^{N} \lambda(i) \overline{F}(G, \mathbf{r}_i')] \;=\; \lambda(i_{ref}) \,[\sum_{g} \{\lambda(i_{ref}) + \sum_{i \neq iref}^{N} \lambda(i) \exp(-ig.\mathbf{r}_i')\} f(g)$$

$$+ F(G).\lambda(i_{ref}) + \sum_{i \neq iref}^{N} \lambda(i) \overline{F}(G, \mathbf{r}_i')] \qquad (x)$$

The co-efficient of f(0) in the first summation appearing in the equation (x) is

$$[\lambda(i_{ref}) + \sum_{i \neq iref}^{N} \lambda(i)] = 0$$

due to the electro-neutrality of the basis. Hence the singularity arising from f (**g**) for **g=0** is removed.